\documentclass[a4paper,11pt]{article}

\usepackage[latin1]{inputenc}

\usepackage{fancyhdr}

\usepackage{graphicx}

\usepackage{amsfonts,amssymb,amscd,latexsym}

\usepackage[intlimits,namelimits,sumlimits]{amsmath}

\usepackage{psfrag}

\renewcommand{\Im}{\mathsf{Im}}
\renewcommand{\Re}{\mathsf{Re}}
\newcommand{\Res}{\mathsf{Res}}
\newcommand{\bra}[1]{\langle #1|}
\newcommand{\ket}[1]{|#1\rangle }
\newcommand{\braket}[2]{\langle #1|#2\rangle}

\newcommand{\eps}{\varepsilon}

\newcommand{\vev}[1]{\langle 0| #1 |0\rangle}
\newcommand{\Fd}{\dot F_{\tau}(\omega)}
\newcommand{\vx}{\mathbf{x}}
\newcommand{\vk}{\mathbf{k}}
\newcommand{\vv}{\mathbf{v}}
\newcommand{\vxi}{\boldsymbol{\xi}}

\newcommand{\vvx}{\mathsf x}
\newcommand{\vvk}{\mathsf k}
\newcommand{\vvu}{\mathsf u}
\newcommand{\vvb}{\mathsf b}
\newcommand{\vve}{\mathsf e}
\newcommand{\lime}{\lim_{\eps\to0}}

\newcommand{\vac}{\ket{0}}
\newcommand{\cav}{\bra{0}}
\newcommand{\infint}{\int_{-\infty}^\infty}
\renewcommand{\d}{\mathrm{d}}

\begin{document}

\title{Considerations on the Unruh Effect:\\ Causality and Regularization}

\author{Sebastian Schlicht    \\
   \mbox{}                                     \\
\normalsize{Fakultät für Physik}           \\
\normalsize{Universität Freiburg}            \\
\normalsize{Hermann-Herder-Straße 3}          \\
\normalsize{D-79104 Freiburg, Germany}         \\
   \mbox{}                                     \\
{\small schlicht@physik.uni-freiburg.de}        \\
}

\date{August 20, 2003}

\maketitle

\begin{abstract}
\noindent
This article is motivated by the observation, that calculations of the Unruh effect
based on idealized particle detectors are usually made in a way that 
involves integrations along the {\em entire} detector trajectory up to the infinitely 
remote {\em future}.
We derive an expression which allows time-dependence of the detector response
in the case of a non-stationary trajectory
and conforms more explicitely to the principle of causality, namely that the response
at a given instant of time depends only on the detectors {\em past} movements.
On trying to reproduce the thermal Unruh spectrum we are led to an unphysical result,
which we trace down to the use of the standard regularization $t\to t-i\eps$ of the
correlation function. 
By consistently employing a rigid detector of finite extension, we are led to a different
regularization which works fine with our causal response function.
\end{abstract}

\section{Introduction}

In 1976, W. G. Unruh \cite{unr} discovered that an uniformly accelerated observer moving 
through {\em empty} Minkowski space sees a thermal radiation with a temperature directly 
proportional to the observers proper acceleration $\alpha$:
\begin{eqnarray*}
T=\frac{\hbar}{2\pi ck}\;\alpha\quad.
\end{eqnarray*}
This surprising and simple result, which blurs the traditional distinction between 
``emptiness'' on the one hand and matter filled space on the other, has given rise to
numerous investigations.

From the many ways of treating the Unruh effect, we concentrate
on the approach originally used by Unruh and refined afterwards by DeWitt \cite{dew}, 
which we consider the most ``down to earth'', namely the concept of an idealized 
particle detector: 
A pointlike quantum mechanical system with different internal 
energy states, which is coupled to a scalar field in its vacuum state and is
moving through Minkowski space along a given trajectory $\vvx(\tau)$. A transition 
of the system from its ground state to an excited state will be interpreted as the
detection of a particle of the corresponding energy.

The reaction of the detector to the motion $\vvx(\tau)$ is formulated via the 
so-called response function $F(\omega)$, which gives in essence the probability of 
finding the detector in an excited state of energy $\omega$ above its ground state.
The usual expression for this response function can be found, for example, in \cite{bd}:
\begin{eqnarray}
\label{eq:bdfomega}
F(\omega)=\infint\d\tau\infint\d\tau'\;e^{-i\omega(\tau-\tau')}\;
\cav\phi(\vvx(\tau))\phi(\vvx(\tau'))\vac\quad.
\end{eqnarray}
It consists of a twofold Fourier transform of the correlation function of the field 
$\phi$, evaluated along the detector trajectory $\vvx(\tau)$. 
It may be considered a bit unfavourable, that this form of the 
response function does express neither time dependence nor causality, by which we mean
the following: It has to be expected, that a particle detector moving through 
Minkowski space along a trajectory which is in general non stationary, will show 
different reactions at different instants of time. It is therefore desirable to have
an expression for its reactions which allows an explicit time dependence, and, what is
more, a time dependence of the kind, that the reaction at a given instant is influenced 
only by the part of the trajectory lying in the {\em past} of this instant. In 
(\ref{eq:bdfomega}) 
there is obviously no room for such a time-dependent and causal behaviour, because
$\tau$ and $\tau'$ are integrated over and $F(\omega)$ can, by construction, never
be time-dependent.

In section 2 we therefore consider the momentary transition rate $\Fd$ instead of
$F(\omega)$. We derive an expression for $\Fd$ which depends in a manifestly causal way
on the 
proper time $\tau$ and which can be applied to non-stationary trajectories. Unfortunately,
the evaluation of this response function leads to an unphysical result already in the 
case of hyperbolic movement.

In section 3 we briefly review parts of an article by S. Takagi \cite{tak}, where he uses 
a correlation function different from the standard one and tries to justify this by 
considering the pointlike Unruh detector as the limit of a rigid and finitely extended
detector. With Takagis' correlation function, our causal response function leads to the
expected result in the Unruh case.

But because we consider Takagis' argumentation not entirely correct, we propose in 
section 4 our own correlation function, which we consider well founded and appropriate
for the problem of the moving detector. This leads on causal calculation to the 
Unruh effect without any difficulties.

In section 5 we show an application to a non-stationary trajectory which smoothly
starts from rest and goes over to an uniform acceleration.

In the final section 6 we briefly state our results and formulate a critical
remark on an often heared argument which relates the thermality of the Unruh radiation 
to the existence of a so-called acceleration horizon.

\bigskip

We use the Minkowski metric with ``spacelike'' signature $(-,+,+,+)$ and denote
three-vectors by boldface,
four-vectors by sans-serif letters. $\hbar$ and $c$ are set equal to 1.

\section{Causal transition rate}

We employ as detector an idealized ``atom'', i.e.\ a quantum mechanical system
with a two-dimensional state space. The Hamiltonian $H_D$ of the
detector atom is of diagonal form with respect to the basis $\ket{0}$ (ground
state) and $\ket{1}$ (excited state). The energy of the ground state is $0$,
the energy of the excited state is denoted by $\omega$. The detector is moving
along the trajectory $t=t(\tau),\;\vx=\vx(\tau)$ through a massless
scalar field $\phi(t,\vx)$ to which it is coupled by the interaction
$\mu(\tau)\phi(\tau)$. $\mu(\tau)$ is the detectors monopole moment,
$\phi(\tau):=\phi(t(\tau),\vx(\tau))$ the scalar field at the detectors
position at time $\tau$. The complete Hamiltonian is $H=H_D+H_F+\mu\phi$, where
the explicit forms of $H_D, H_F$ and $\mu$ can be left unspecified.

We now ask the following question: Suppose an energy measurement at time
$\tau_0$ has given the value $0$ for the detector atom and the vacuum state for
the scalar field. Then, what is the probability of finding the excited state of
the detector, i.e.\ the energy value $\omega$, in a later measurement at time
$\tau>\tau_0$ if the detector has in the meantime been moving along a given
trajectory?
We do the calculation in the interaction picture, in which both, states and operators,
are time-dependent. The field operator has the form
\begin{eqnarray}
\label{scf}
\phi(t,\vx)=\frac{1}{(2\pi)^3}\int\frac{\d^3k}{2\omega(\vk)}\;
\left(a(\vk)e^{-i(\omega(\vk)t-\vk\vx)}+a^\dag(\vk)e^{i(\omega(\vk)t-\vk\vx)}\right)
\end{eqnarray}
where the creation- and annihilation-operators obey the covariant
commutation relations
\begin{eqnarray*}
[a(\vk),a^\dag(\vk')]=(2\pi)^3 2\omega(\vk)\delta(\vk-\vk')\quad,
\end{eqnarray*}
and $\omega(\vk)=|\vk|$. The time evolution of the states is governed by
\begin{eqnarray*}
i\frac{\d}{\d\tau}\ket{\psi(\tau)}=\mu(\tau)\phi(\tau)\ket{\psi(\tau)}
\end{eqnarray*}
where
\begin{eqnarray*}
\mu(\tau)=e^{iH_D\tau}\mu(0)e^{-iH_D\tau}\quad.
\end{eqnarray*}
At $\tau_0$ the system is in the product state $\ket{\psi(\tau_0)}=\vac\vac$,
i.e.\ both the detector and the field are in their ground state. Then to the first
order of perturbation theory
\begin{eqnarray*}
|\braket{1\varphi}{\psi(\tau)}|^2=|\bra{1}\mu(0)\ket{0}|^2
\int_{\tau_0}^\tau\d\tau'\int_{\tau_0}^\tau\d\tau''\;e^{-i\omega(\tau'-\tau'')}
\cav\phi(\tau')\ket{\varphi}\bra{\varphi}\phi(\tau'')\vac
\end{eqnarray*}
is the probability to find the detector in the state $\ket{1}$ and the
field in the state $\ket{\varphi}$ at time $\tau$. Summing over the unobserved 
field states and using their completeness, we arrive at
\begin{eqnarray*}
\sum_{\ket{\varphi}}|\braket{1\varphi}{\psi(\tau)}|^2=|\bra{1}\mu(0)\ket{0}|^2
\int_{\tau_0}^\tau\d\tau'\int_{\tau_0}^\tau\d\tau''\;e^{-i\omega(\tau'-\tau'')}
\cav\phi(\tau')\phi(\tau'')\vac\quad,
\end{eqnarray*}
which is the desired probability of finding the detector in its excited state at 
time $\tau$. It consists of a constant, detector-specific coefficient and an integral
which depends only on the field $\phi$. We are interested only in the integral
which we denote by $F_\tau(\omega)$, and we introduce the abbreviation 
$W(\tau,\tau')$ (``Wightman-function'') for the correlation function 
$\cav\phi(\tau)\phi( \tau')\vac$:
\begin{eqnarray*}
F_\tau(\omega)=\int_{\tau_0}^\tau\d\tau'\int_{\tau_0}^\tau\d\tau''\;
e^{-i\omega(\tau'-\tau'')}W(\tau,\tau')\quad.
\end{eqnarray*}
The integration extends over a square in the $\tau',\tau''$-plane. 
We can write this in a more convenient form by introducing new
coordinates $u=\tau'\;,\; s=\tau'-\tau''$ in the lower triangle $\tau''<\tau$,
and $u=\tau''\;,\; s=\tau''-\tau'$ in the upper triangle $\tau''>\tau'$.
This leads to
\begin{eqnarray*}
F_{\tau}(\omega)= \int_{\tau_0}^{\tau}\d u\int_0^{u-\tau_0}\d s\; \left( 
e^{-i\omega s}\; W(u,u-s)\;+\;e^{+i\omega s}\; W(u-s,u)\right)\quad,
\end{eqnarray*}
which is the same as
\begin{eqnarray*}
F_{\tau}(\omega)=2\int_{\tau_0}^{\tau}\d u\int_0^{u-\tau_0}\d s\;\Re\left(
e^{-i\omega s}\;W(u,u-s)\right)
\end{eqnarray*}
because of $W(\tau',\tau)=W^*(\tau,\tau')$.
We can read off the following structure: The probability of finding the detector at 
time $\tau$ in a state, whose energy is different by an amount $\omega$ from the 
energy of the ground state, is given as a sum of contributions from every time $u$
between $\tau_0$ and $\tau$. The contribution of a certain time $u$ is in turn given
by the correlations of $\phi(u)$ with $\phi$ at every past time since $\tau_0$.
Therefore, the time derivative of $F_{\tau}(\omega)$, i.e.\ the transition rate $\Fd$,
is not determined only by the state of motion at time $\tau$ but by the whole course of
the trajectory between $\tau_0$ and $\tau$:
\begin{eqnarray*}
\Fd=2\int_0^{\tau-\tau_0}\d s\;\Re\left(e^{-i\omega s}\;W(\tau,\tau-s)\right)\quad.
\end{eqnarray*}
From now on we will consider only the transition rate $\Fd$ which is proportional to
the number of ``clicks'' per second in a detector consisting of a swarm of identical
detector atoms, and which is physically
more accessible than the probability $F_{\tau}(\omega)$. (In the following we will
call $\Fd$ ``response-function'', which normally rather is the name for the probability
$F_\tau (\omega)$.)
Furthermore we get rid
of the arbitrariliy chosen $\tau_0$ by doing the limit $\tau_0\to-\infty$:
\begin{eqnarray}
\label{eq:Fd}
\Fd=2\int_0^\infty\d s\;\Re\left(e^{-i\omega s}\;W(\tau,\tau-s)\right)\quad.
\end{eqnarray}
If $W$ is invariant under time translation, i.e.\
$W(\tau_1+\Delta\tau,\tau_2+\Delta\tau)=W(\tau_1,\tau_2)$ then it is an easy matter
to show that 
\begin{eqnarray}
\label{eq:ttinv}
&&2\int^{\infty}_0 \d s\; \Re\left( e^{-i\omega s}\; 
W(\tau,\tau-s)\right)
=\int_{-\infty}^{\infty}\d s\;e^{-i\omega s}\;W(s,0)\quad.
\end{eqnarray}
That means, if $W$ is invariant under time translation, $\Fd$ is 
independent of $\tau$ and can formally be written as 
an integral over the {\em whole} (i.e.\ past and {\em future}) trajectory.
This does not lead to any problem in the case of a stationary trajectory, because
this is a ``global'' object anyway, for
an arbitrarily small piece of such a trajectory contains all the information 
about its whole, in particular its future, course.
But in the general case of a non-stationary trajectory, one must 
use (\ref{eq:Fd}) which is therefore the starting point of our further considerations.

\bigskip

Up to now, our reasoning was completely general and we did not make any use of the fact 
that $\phi$ is a scalar field of the form (\ref{scf}). 
In the infinite-dimensional case of the scalar field, it has to be taken into account 
that $W$ is not a function but a distribution. This is clear from 
\begin{eqnarray*}
\cav\phi(\vvx)\phi(\vvx')\vac&=&\frac{1}{(2\pi)^6}\;\cav
\int\frac{\d^3k\;\d^3k'}{4\omega(\vk) 
\omega(\vk')}\;a(\vk) a^\dag(\vk')\; e^{i(\vvk\vvx-\vvk'\vvx')}\;\vac\\
&=&\frac{1}{(2\pi)^3}\int\frac{\d^3k}{2\omega(\vk)}\;e^{i\vvk(\vvx-\vvx')}\quad,
\end{eqnarray*}
because the integrand is oscillatory and the integral does not converge in the 
classical sense. The standard way to handle this is to 
make the replacement $t\to t-i\eps$, or, equivalent, to insert an exponential factor
$e^{-|\vk|\eps}$ into the Fourier representation, which regularizes the 
high-frequency behavior and take the limit $\eps\to 0$ at the end of the calculation,
i.e.
\begin{eqnarray}
\label{eq:cfreg}
\cav\phi(\vvx)\phi(\vvx')\vac=
\frac{1}{(2\pi)^3}\int\frac{\d^3k}{2\omega(\vk)}\;e^{i\vvk(\vvx-\vvx')-|\vk|\eps}
\end{eqnarray}
and
\begin{eqnarray}
\label{eq:Fdreg}
\Fd=2\lime\int_0^\infty\d s\;\Re\left(e^{-i\omega s}\;W_\eps(\tau,\tau-s)\right)\quad.
\end{eqnarray}
We will investigate later how this regularization can be done 
in a more physically motivated way. Accepting the $e^{-|\vk|\eps}$-factor for the 
moment, one arrives after a short calculation at
\begin{eqnarray}
\label{eq:bdcf}
\cav\phi(\vvx)\phi(\vvx')\vac=\frac{-1/4\pi^2}{(t-t'-i\eps)^2-(\vx-\vx')^2}\quad.
\end{eqnarray}
This is the form of the correlation function which is used in the standard literature, 
see e.g.\ \cite{bd}, (3.59). If the trajectory of a detector at rest, 
$t(\tau)=\tau,\;\vx(\tau)=0$ is inserted, one gets 
\begin{eqnarray*}
\Fd=-\frac{1}{2\pi^2}\lime\int_0^\infty\d s\;\Re\;\frac{e^{-i\omega s}}{(s-i\eps)^2}=
-\frac{1}{4\pi^2}\lime\int_{-\infty}^\infty\d s\;\frac{e^{-i\omega s}}{(s-i\eps)^2}\quad.
\end{eqnarray*}
One can evaluate this integral by contour-integration along the real axis closed by an
infinite semi-circle in the upper or lower half-plane, depending on the sign of $\omega$.
The resulting value is $2\pi\omega e^{\eps\omega}\Theta(-\omega)$ which leads to
\begin{eqnarray*}
\Fd=-\frac{1}{2\pi}\;\omega\Theta(-\omega)\quad.
\end{eqnarray*}
This result - independent of $\tau$ as expected - describes the fact that a detector atom
at rest does not go spontaneously into its excited state ($\Fd=0\;\forall\omega>0$), but
of course it decays spontaneously into its ground state due to the coupling to the 
scalar field.

\bigskip

\begin{figure}[htbp]
\begin{center}
\psfrag{om}[][][0.7]{$\omega$}
\psfrag{Fd}[][][0.7]{$\Fd$}
\includegraphics[width=11cm]{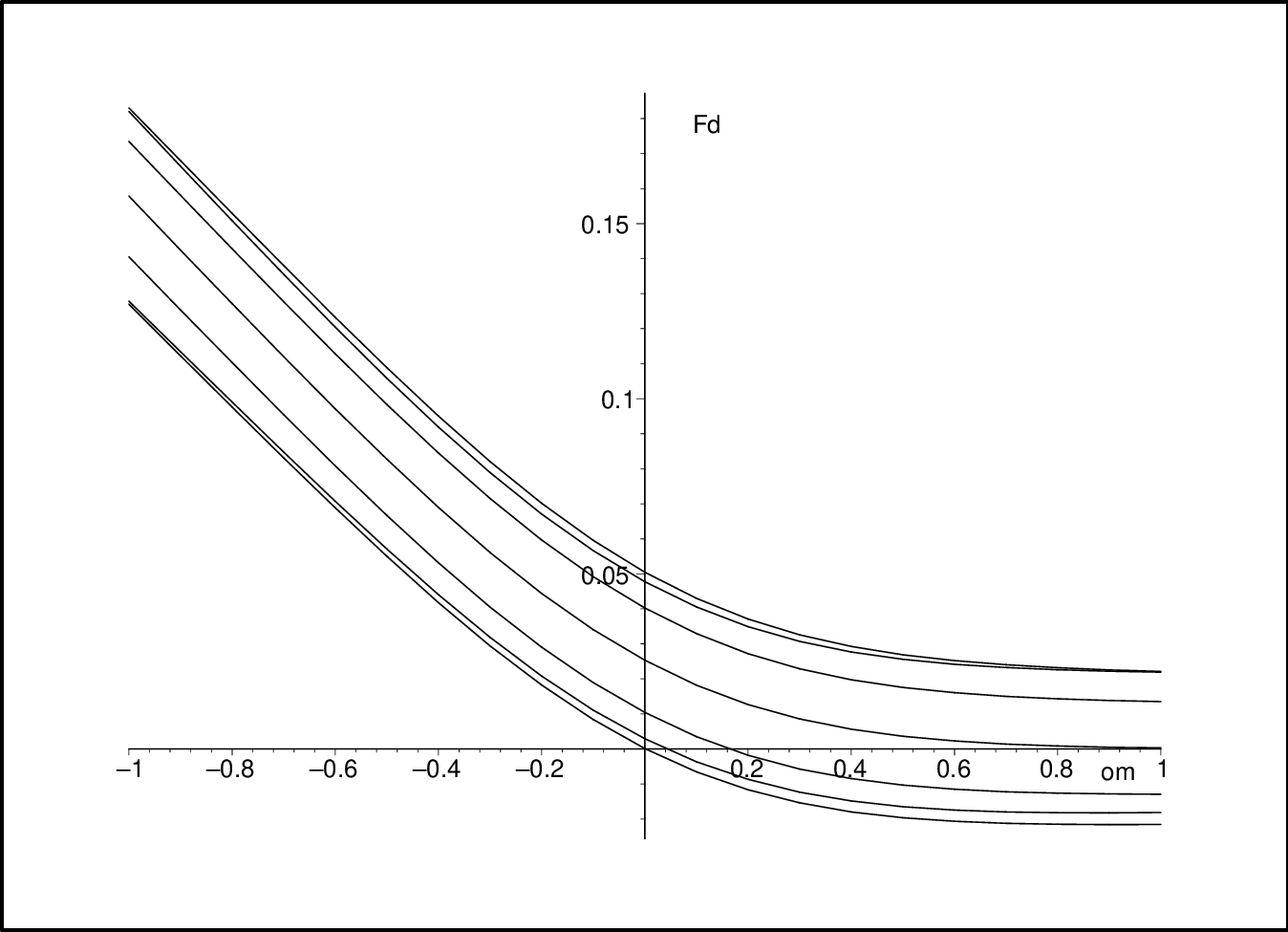}
\caption{Numerical results for $\Fd$}\label{num}
\end{center}
\end{figure}

Now the interesting thing is to see what happens if the detector is uniformly accelerated,
i.e.\footnote{We set the proper acceleration equal to 1 because the transition-rate for
arbitrary acceleration $\alpha>0$ can be obtained by a simple scale-transformation 
$\Fd=\alpha\dot{F}_{\alpha\tau}^{(\alpha=1)}\left(\frac{\omega}{\alpha}\right)$.}
\begin{eqnarray*}
t(\tau)=\sinh(\tau),\quad x(\tau)=\cosh(\tau),\quad
y(\tau)=z(\tau)=0\quad.
\end{eqnarray*}
Because this is an orbit of the Killing vector $x\partial_t+t\partial_x$ it is stationary
and one expects a time-independent transition rate. Therefore it is a bit surprising 
to see that insertion of this trajectory into (\ref{eq:bdcf}) leads to a $W_\eps(\tau,\tau-s)$ 
which is {\em not} time-independent, i.e.\ not independent of $\tau$:
\begin{eqnarray*}
W_\eps(\tau,\tau-s)=\frac{-1/4\pi^2}{4\sinh^2\left(\frac{s}{2}\right)
-\eps^2-2i\eps\left(\sinh(\tau)-\sinh(\tau-s)\right)}\quad.
\end{eqnarray*}
But this in itself is not to alarming, because time independence is demanded only for
the transition rate, which contains a limit $\eps\to 0$, and $W_0(\tau,\tau-s)$ 
is clearly independent of $\tau$. To settle the case, let us do a numerical calculation
of $\Fd$:
The response function (\ref{eq:Fdreg}) is of the form
\begin{eqnarray*}
\Fd&=&2\lime\int_0^\infty\d s\;\left(a_\eps(\tau,\tau-s)\cos\omega s + 
b_\eps(\tau,\tau-s)\sin\omega s\right)\quad,
\end{eqnarray*}
where $a_\eps:=\Re W_\eps$ and $b_\eps:=\Im W_\eps$, i.e.
\begin{eqnarray*}
a_{\eps}(\tau,\tau-s)&=&-\frac{1}{4\pi^2}\frac{4\sinh^2(\frac{s}{2})-
\eps^2}{\left(4\sinh^2(\frac{s}{2})-\eps^2\right)^2+
4\eps^2(\sinh\tau-\sinh(\tau-s))^2}\\
b_{\eps}(\tau,\tau-s)&=&-\frac{1}{4\pi^2}\frac{2\eps(\sinh\tau-
\sinh(\tau-s))}{\left(4\sinh^2(\frac{s}{2})-\eps^2\right)^2+
4\eps^2(\sinh\tau-\sinh(\tau-s))^2}\quad.
\end{eqnarray*}
In order to calculate the transition rate for a given time $\tau$, we thus have to 
evaluate the cosine- or sine-transform of $a_\eps(\tau,\tau-s)$ and $b_\eps(\tau,\tau-s)$
with respect to $s$ at fixed $\tau$. 
The numerical results for seven different times $\tau=0,\pm1,\pm2,\pm4$ are shown in 
figure 
\ref{num}. The lower curve corresponds to $\tau=-4$, higher curves to 
increasingly later times.
Thus our suspicion is confirmed that the $\tau$-dependence of the correlation function
leads to a $\tau$-dependent transition rate, which is unacceptable in the case at hand.
Furthermore, the results are problematic for $\tau<0$ because $\Fd$ assumes negative
values in this cases.

At this point, the first question which comes to mind is to ask the reliability
of the numerical scheme. Fortunately, an exact 
calculation is possible at least in the two special cases of $\tau=0$ or $\omega=0$.
In the case $\tau=0$ the transition rate can be evaluated by the method of 
contour-integration. This results in
\begin{eqnarray*}
\dot{F}_0(\omega)=\frac{1}{2\pi}\frac{\omega}{e^{2\pi\omega}-1}\quad,
\end{eqnarray*}
which is indeed the usual Unruh spectrum. This is exactly in agreement with the 
numerical result for $\tau=0$.
In the other case, $\omega=0$, the integral can be evaluated exactly with the help of
Maple, which leads to
\begin{eqnarray*}
\dot{F}_\tau (0)=\frac{1+e^{2\tau}(2\tau-1)}{2\pi^2(e^{2\tau}-1)^2}\quad.
\end{eqnarray*}
In this way we get on the one hand a rigorous proof that $\Fd$ is indeed $\tau$-dependent
if the calculation is based on (\ref{eq:bdcf}).
On the other hand this exact result permits a further test of the numerical calculation. 
For example, the numerics tells us that $\dot{F}_1(0)=0.0104...$, and indeed
\begin{eqnarray*}
\dot{F}_1(0)=\frac{1+e^2}{2\pi^2(e^2-1)^2}=0.0104...\quad.
\end{eqnarray*}
For any other checked $\dot{F}_\tau (0)$ there is an equally good agreement between the 
numerical prediction and the analytic result. This gives us confidence in the 
correctness of the numerical calculation in the general case. 

\bigskip

We are thus confronted with the following situation: The application of our causal
formulation of the detector response to the case of uniformly accelerated motion 
lead to a unphysical, time-dependent result. But we see no 
reason to believe that it is impossible to derive the Unruh effect consistently
in an explicitly causal way, or, more generally, to formulate a causal response-function
which can be applied to non-stationary trajectories. Therefore, we will seek the error
not in the use of our causal response function (\ref{eq:Fdreg}), but rather suspect 
that the 
correlation function (\ref{eq:bdcf}) and the $e^{-|\vk|\eps}$-regularization 
of the integral representation (\ref{eq:cfreg}) is not suitable for the problem of the 
moving 
detector. In the following sections we will consider a finite, rigid detector and show
that one is led in a natural way to a correlation function which is {\em different} 
from (\ref{eq:bdcf})\footnote{In \cite{bd} the Unruh spectrum is derived from 
(\ref{eq:bdcf}) only by the use of a not entirely convincing calculation in which
"a positive function of $\tau,\tau'$ is absorbed into $\eps$".}. 
We use the article \cite{tak} of S. Takagi as a model, even if we 
come to 
a different result. The new correlation function leads upon causal calculation to the 
desired Unruh result without any difficulty, and it seems to be superior to 
(\ref{eq:bdcf}) in other ways.

\section{Correlation function a la Takagi}

In \cite{tak} Takagi criticizes the ad-hoc regularization of the correlation function
(\ref{eq:cfreg}) by the introduction of the factor $e^{-|\vk|\eps}$ into its 
Fourier representation
or by the equivalent prescription $t\to t-i\eps$. Rather he would like to understand
the origin of the regulator as a consequence of the finite extension of the detector.
Therefore he uses instead of the ultralocal $\phi(\tau)=\phi(\vvx(\tau))$ the 
``smeared'' version
\begin{eqnarray*}
\phi(\tau)=\int\d^3\xi\;f(\vxi)\;\phi(\vvx(\tau,\vxi))\quad,
\end{eqnarray*}
where $f(\vxi)$ is a weight function which is normalized, symmetric under rotations
and concentrated around the origin. By choosing $f(\vxi)=\delta(\vxi)$ one recovers
the pointlike detector; we will consider a family of finite detectors 
$f_\eps(\vxi)$ instead with $\lime f_\eps(\vxi)=\delta(\vxi)$. By taking the limit
only at the end of the calculation, we thus model an ``infinitesimal'' instead 
of a pointlike detector. 

Of special importance is the function $\vvx(\tau,\vxi)$, so we have to say a little
bit about it.
It describes the transformation from the usual Minkowski coordinates to the
Fermi-Walker coordinates $\tau$ and $\vxi:=(\xi,\eta,\zeta)$ associated with the 
trajectory\footnote{For the moment, we restrict ourselves to trajectories which run along 
the $x$-axis, i.e.\ $y=z=0$} $\vvx=\vvx(\tau)$:
\begin{eqnarray}
\label{eq:fko}
t(\tau,\vxi)&=&\quad\; t(\tau)+\xi{\dot x}(\tau)\nonumber \\
\vx(\tau,\vxi)&=&\left( \begin{array}{c} x(\tau)+\xi{\dot t}(\tau)\\
\eta \\ \zeta 
\end{array} \right)\quad.
\end{eqnarray}
The geometrical meaning of these coordinates is the following: To every point 
$\vvx(\tau)=(t(\tau),x(\tau),0,0)$ on the trajectory belongs its simultaneity,
i.e.\ the hyperplane orthogonal to the four-velocity 
$\vvu(\tau)=(\dot{t}(\tau),\dot{x}(\tau),0,0)$ in this point. This three-dimensional
space consists of all events which are simultaneous to $\vvx(\tau)$, where simultanity
is judged from the co-moving inertial frame. If we attach orthonormal basis vectors 
$\vve_{\xi}(\tau)=({\dot x}(\tau),{\dot t}(\tau),0,0)\;,\;
\vve_{\eta}(\tau)=(0,0,1,0)\;,\;\vve_{\zeta}(\tau)=(0,0,0,1)$ to every such hyperplane,
we can characterize every event $\vvx$ (in a neighborhood of the trajectory) by
$(\tau,\xi,\eta,\zeta)$: First, we fix $\tau$ by the condition that the given 
spacetime point lies in the simultanity of the trajectory point $\vvx(\tau)$. 
Then $\vvx-\vvx(\tau)$ is a linear combination of $\vve_{\xi}(\tau),\vve_{\eta}(\tau),
\vve_{\zeta}(\tau)$, and we can read off $(\xi,\eta,\zeta)$ as the corresponding
coefficients. That is\footnote{$\tau$ is used with two different meanings: As proper time 
of the trajectory, and as spacetime coordinate. This should not lead to any confusion.} 
\begin{eqnarray*}
\vvx(\tau)+\xi\vve_{\xi}(\tau)+\eta\vve_{\eta}(\tau)+\zeta\vve_{\zeta}(\tau)=
t\vve_t+x\vve_x+y\vve_y+z\vve_z\quad,
\end{eqnarray*}
which is nothing else but (\ref{eq:fko}). 
Expressed by the coordinates $(\tau,\vxi)$, the Minkowski metric assumes the form
\begin{eqnarray*}
\d s^2=-\left(1\,+\,2({\dot t}{\ddot x}-{\dot x}{\ddot t})\xi
\,+\,({\ddot x}^2-{\ddot t}^2)\xi^2\right)\d\tau^2+\d\vxi^2\quad.
\end{eqnarray*}
Obviously, every simultanity $\tau=\mathrm{const.}$ carries the same cartesian 
three-metric.
Because $\phi(\tau)$ originates from an integration over the $\tau$-simultanity, 
the $\tau$-independent shape function $f(\vxi)$ ensures that this construction 
corresponds indeed to a {\em rigid} detector - where rigidity means that its 
three-geometry as seen from its own momentary rest system is unchanged in the course of 
proper time. In the case of hyperbolic motion we have for example
\begin{eqnarray*}
t(\tau,\vxi)&=&\quad\; (1+\xi)\sinh\tau\nonumber \\
\vx(\tau,\vxi)&=&\left( \begin{array}{c} (1+\xi)\cosh\tau\\
\eta \\ \zeta 
\end{array} \right)\quad,
\end{eqnarray*}
and the metric 
\begin{eqnarray*}
\d s^2=-(1+\xi)^2\d\tau^2+\d\vxi^2\quad,
\end{eqnarray*}
i.e.\ the well known transformation to Rindler coordinates, cf. \cite{rind}.

Now by linearizing $\vvx(\tau,\vxi)$ around $\vxi=0$
\begin{eqnarray*}
\vvx(\tau,\vxi)\approx\vvx(\tau)\,+\,\vxi\,\frac{\partial\vvx(\tau,\vxi)}
{\partial\vxi}\Big{|}_{\vxi=0}\;=:\;\vvx(\tau)\,+\,\vxi{\sf X}_{\vxi}(\tau)
\end{eqnarray*}
Takagi finds
\begin{eqnarray*}
\phi(\tau)=\frac{1}{(2\pi)^3}\int \frac{\d^3k}{2|\vk|}\;
\left( a(\vk)e^{-i\vvk\vvx(\tau)}{\tilde f}(\vk;\tau)\; + 
\;a^\dag(\vk)e^{i\vvk\vvx(\tau)}{\tilde f}^*(\vk;\tau)\right)\quad,
\end{eqnarray*}
where
\begin{eqnarray*}
{\tilde f}(\vk;\tau):=\int \d^3\xi\;f(\vxi)\;
e^{-i\vvk\,\vxi{\sf X}_{\vxi}(\tau)}
\end{eqnarray*}
converges to $0$ for $|\vk|\to\infty$, which makes $\phi(\tau)$ a well-defined operator.
He now chooses for $f(\vxi)$ the family of functions
\begin{eqnarray*}
f_{\eps}(\vxi)=\frac{1}{\pi^3}\frac{\eps^3}
{(\xi^2+\eps^2)(\eta^2+\eps^2)(\zeta^2+\eps^2)}\quad,
\end{eqnarray*}
which converges to the three-dimensional $\delta$-function in the limit $\eps\to0$.
Then ${\tilde f}(\vk;\tau)$ can be calculated explicitly:
\begin{eqnarray}
\label{eq:feig}
{\tilde f}_{\eps}(\vk;\tau)=e^{-\eps|\vvk{\sf X}_{\xi}(\tau)|-
\eps|\vvk{\sf X}_{\eta}(\tau)|-\eps|\vvk{\sf X}_{\zeta}(\tau)|}\quad.
\end{eqnarray}
Takagi now claims that this complicated expression can be effectively replaced by the
simpler
\begin{eqnarray}
\label{eq:fein}
{\tilde f}_{\eps}(\vk;\tau)=e^{-|\vk|\eps/2}\quad,
\end{eqnarray}
which does not depend on $\tau$ any more. Accepting this for the moment, one arrives
after a few lines at
\begin{eqnarray}
\label{eq:tcf}
\cav\phi(\tau)\phi(\tau')\vac=
\frac{1}{(2\pi)^3}\int\frac{\d^3k}{2|\vk|}e^{i\vvk(\vvx-\vvx')-
|\vk|\eps}\quad.
\end{eqnarray}
Obviously, the finite extension $\eps>0$ of the detector manifests itself in a 
regularization of the integral, as desired.
Takagi now writes this in four-dimensional form
\begin{eqnarray}
\label{eq:tcf4d}
\cav\phi(\tau)\phi(\tau')\vac=\frac{1}{(2\pi)^3}\int \d^4k\;\Theta(k^0)
\delta(\vvk^2)e^{i\vvk(\vvx-\vvx')-k^0\eps}\quad,
\end{eqnarray}
and claims this form to be manifestly Lorentz invariant.
He now uses this Lorentz invariance to derive
\begin{eqnarray}
\label{eq:V}
\cav\phi(\tau)\phi(\tau')\vac=\frac{-1/4\pi^2}{\left({\rm sgn}(t-t')
\sqrt{-(\vvx-\vvx')^2}-i\eps\right)^2}\quad.
\end{eqnarray}
By inserting the hyperbolic trajectory and using contour-integration, it follows
indeed that
\begin{eqnarray*}
\Fd=\frac{1}{2\pi}\frac{\omega}{e^{2\pi\omega}-1}\quad.
\end{eqnarray*}

Even if this is the desired result, we have some reservations concerning the derivation.
First, the claim that the integral (\ref{eq:tcf4d}) is Lorentz invariant holds only if
$\eps=0$. But in our case $\eps>0$ the factor $e^{-k^0\eps}$ destroys this invariance,
which makes the evaluation of the integral obsolete. Indeed, straightforward evaluation 
of (\ref{eq:tcf}) leads exactly to the standard form (\ref{eq:bdcf}) of the 
correlation function and not to (\ref{eq:V}). 
Therefore, if Takagi's derivation were correct up to (\ref{eq:tcf}) this would show that 
taking 
into account the finite extension of the detector does not solve the problem which we
encountered in the last section. We now suppose, that, contrary to Takagi's assertion,
it is {\em not} allowed to replace (\ref{eq:feig}) by (\ref{eq:fein}). 
Indeed this replacement eliminates
the dependence on $\tau$ and on the trajectory, and it is not clear that this leaves
the final result unchanged. Therefore we will try to work out Takagi's idea more 
carefully in the next section.

\section{Correlation function of an infinitesimal rigid detector}

We now use the monopole interaction with a ``smeared'' field-operator

\begin{equation}
\phi(\tau)=\int\d^3\xi\;f(\vxi)\;\phi(\vvx(\tau,\vxi))
\end{equation}
instead of $\phi(\tau)=\phi(\vvx(\tau))$. The correlation function thus assumes
the form:
\begin{eqnarray*}
\cav\phi(\tau)\phi(\tau')\vac&=&\frac{1}{(2\pi)^6}\;\cav\int\d^3\xi\;f(\vxi)
\int\d^3\xi'\;f(\vxi')\\
&&\int\frac{\d^3k\;\d^3k'}{4\omega(\vk)\omega(\vk')}a(\vk)a^\dag(\vk')\;
e^{i\vvk\vvx(\tau,\vxi)}e^{-i\vvk'\vvx(\tau',\vxi')}\;\vac\quad.
\end{eqnarray*}
Using the commutation relations of the
$a$ and $a^\dag$, $a\vac=0$ and $\braket{0}{0}=1$ we can do the
$\vk'$-integration and find \begin{eqnarray*}
\cav\phi(\tau)\phi(\tau')\vac&=&\frac{1}{(2\pi)^3}\int\frac{\d^3k}
{2\omega(\vk)}\;g(\vk;\tau)g^*(\vk;\tau') \end{eqnarray*} where we have introduced
$g(\vk;\tau):=\int\d^3\xi\;f(\vxi)e^{i\vvk\vvx(\tau,\vxi)}$.
In order to find the correlation function of a detector with shape $f(\vxi)$
and trajectory $\vvx(\tau)$ we thus have to determine the transformation to 
Fermi coordinates $\vvx(\tau,\vxi)$, from that the function $g(\vk;\tau)$, and
finally $\cav\phi(\tau)\phi(\tau')\vac$. This amounts to six integrations, so one
would expect difficult computations for even the simplest trajectories. In fact we will
see, that by choosing a suitable shape $f(\vxi)$ of the detector, one arrives at an 
explicit form of the correlation-function for any arbitrary trajectory.

Let us first consider the simple example of a detector which is not moving at all, and 
whose shape is given by a function from the family
\begin{eqnarray}
\label{eq:det}
f_\eps(\vxi)=\frac{1}{\pi^2}\frac{\eps}{(\vxi^2+\eps^2)^2}\quad,\quad\eps>0\quad.
\end{eqnarray}
The $f_\eps(\vxi)$ are invariant under rotations, satisfy 
$\int\d^3\xi\;f_\eps(\vxi)=1\quad\forall\eps>0$ and have the scaling property
$f_\eps(\vxi)=\eps^{-3}f_1(\vxi/\eps)$. This means, they approximate the three-dimensional
$\delta$-function. The parameter $\eps$ has the physical dimension of a length and 
indicates the extension of the detector: $f_\eps(\vxi)$ is negligible outside
$\vxi^2<\eps^2$. By taking the limit $ \eps\to 0$ at the end of the computation
one can model an infinitesimal detector. The transformation to Fermi coordinates is
trivial in the case of a detector at rest: $t(\tau,\vxi)=\tau,\;\vx(\tau,\vxi)=\vxi$.
This leads to
\begin{eqnarray*}
g_\eps(\vk;\tau)=\int\d^3\xi\;f_\eps(\vxi)e^{i\vvk\vvx(\tau,\vxi)}=
e^{-i|\vk|\tau}e^{-\eps|\vk|}\quad.
\end{eqnarray*}
From that:
\begin{eqnarray}
\label{eq:ruhe}
\cav\phi(\tau)\phi(\tau')\vac&=&\frac{1}{(2\pi)^3}\int\frac{\d^3 k}{2|\vk|}\;
g_\eps(\vk;\tau)g_\eps^*(\vk;\tau')\nonumber\\
&=&\frac{1}{(2\pi)^3}\int\d^3 k\;\frac{e^{-i|\vk|(\tau-\tau'-i\eps)}}{2|\vk|}\nonumber\\
&=&\frac{-1/4\pi^2}{(\tau-\tau'-i\eps)^2}\quad.
\end{eqnarray}
(We have replaced $2\eps$ by $\eps$.) Using the same contour-integration as before in 
section 2 it follows 
\begin{eqnarray*}
\Fd=-\frac{1}{2\pi}\omega\Theta(-\omega)\quad.
\end{eqnarray*}
So in this simple case we get the same result as before in section 2 where we
inserted the trajectory into the Wightman function (\ref{eq:bdcf}). But we now see that
the high-frequency regulator $e^{-\eps|\vk|}$, which was put in by hand in section 2,
results from the finite extension of the rigid detector and thus acquires a
physically meaningful interpretation. 

\bigskip

Let us now consider an arbitrary trajectory $\vvx=\vvx(\tau)$. The transformation
to Fermi coordinates is
\begin{eqnarray*}
\vvx(\tau,\vxi)=\vvx(\tau)+\xi\vve_\xi(\tau)+\eta\vve_\eta(\tau)+\zeta\vve_\zeta(\tau)
\quad,
\end{eqnarray*}
where $\vve_\xi(\tau),\vve_\eta(\tau),\vve_\zeta(\tau)$ together with 
$\vvu(\tau)=\dot\vvx(\tau)$ form an orthonormal basis which is Fermi-Walker transported
along $\vvx(\tau)$. We will use the same detector-shape (\ref{eq:det}) as in the case of 
the detector at rest. Let us start with the calculation of $g_\eps(\vk;\tau)$:
\begin{eqnarray*}
g_\eps(\vk;\tau)=\int\d^3\xi\;f_\eps(\vxi)e^{i\vvk\vvx(\tau,\vxi)}=
e^{i\vvk\vvx(\tau)}\int\d^3\xi\;f_\eps(\vxi)
e^{i(\xi\vvk\vve_\xi(\tau)+\eta\vvk\vve_\eta(\tau)+\zeta\vvk\vve_\zeta(\tau))}\quad.
\end{eqnarray*}
We now introduce the abbreviation
$\bar\vk:=(\vvk\vve_\xi(\tau),\vvk\vve_\eta(\tau),\vvk\vve_\zeta(\tau))$. 
It is easy to do the integration by using spherical coordinates in $\vxi$-space:
\begin{eqnarray*}
\int\d^3\xi\;f_\eps(\vxi)\;e^{i\bar\vk\vxi}=
e^{-\eps|\bar\vk|}\quad.
\end{eqnarray*}
That means
\begin{eqnarray*}
g_\eps(\vk;\tau)=e^{i\vvk\vvx(\tau)}e^{-\eps\sqrt
{(\vvk\vve_\xi(\tau))^2+(\vvk\vve_\eta(\tau))^2+(\vvk\vve_\zeta(\tau))^2}}\quad.
\end{eqnarray*}
Because $\vvk$ is a lightlike four-vector, the sum of squares under the 
square root is a complete square: By multiplying the identity
\begin{eqnarray*}
\vvk=-(\vvk\vvu)\vvu+(\vvk\vve_\xi)\vve_\xi+(\vvk\vve_\eta)\vve_\eta
+(\vvk\vve_\zeta)\vve_\zeta
\end{eqnarray*}
with $\vvk$ we get
\begin{eqnarray*}
(\vvk\vve_\xi(\tau))^2+(\vvk\vve_\eta(\tau))^2+(\vvk\vve_\zeta(\tau))^2=
(\vvk\vvu)^2=(-|\vk|\dot t+\vk\dot\vx)^2\quad.
\end{eqnarray*}
$\vvk\vvu$ is the product of a lightlike and a timelike vector and therefore negative,
so that
\begin{eqnarray*}
g_\eps(\vk;\tau)=e^{i\vvk\vvx(\tau)}e^{+\eps\vvk\vvu(\tau)}\quad.
\end{eqnarray*}
Obviously, $e^{\eps\vvk\vvu(\tau)}$ and not simply $e^{-\eps|\vk|}$, is the 
physically motivated regularisating factor resulting from the finite extension of
the detector.
Using this, we now calculate
\begin{eqnarray*}
\int\frac{\d^3k}{|\vk|}\;g_{\eps}(\vk;\tau)g_{\eps}^*(\vk;\tau')&=&
\int\frac{\d^3k}{|\vk|}\;e^{i\vvk(\vvx-\vvx')}\,e^{+\eps\vvk(\vvu+\vvu')}\\
&=&\int \d^3k\;e^{i\vk(\vx-\vx'-i\eps({\dot \vx}+{\dot \vx}'))} 
\;\frac{e^{-i|\vk|(t-t'-i\eps({\dot t}+{\dot t}'))}}{|\vk|}\\
&=&\frac{4\pi}{-(t-t'-i\eps({\dot t}+{\dot t}'))^2+
(\vx-\vx'-i\eps({\dot \vx}+{\dot \vx}'))^2}\\
+&=&\frac{4\pi}{\left(\vvx-\vvx'-i\eps({\dot \vvx}+{\dot \vvx}')\right)^2}
\quad.
\end{eqnarray*}
From this we finally obtain
\begin{eqnarray}
\label{eq:meinW}
\cav\phi(\tau)\phi(\tau')\vac=
\frac{1/4\pi^2}{\bigl( 
\vvx(\tau)-\vvx(\tau') -i\eps({\dot\vvx}(\tau)+
{\dot\vvx}(\tau'))\bigr)^2}\quad.
\end{eqnarray}
This result has to be compared with
\begin{eqnarray}
\label{eq:bdW}
\cav\phi(\vvx(\tau))\phi(\vvx(\tau'))\vac=\frac{-1/4\pi^2}{\left(t(\tau)
-t(\tau')-i\eps\right)^2 - \left(\vx(\tau)-\vx(\tau')\right)^2}\quad,
\end{eqnarray}
which is obtained from the usual Wightman function (\ref{eq:bdcf}) by inserting the 
trajectory 
$\vvx=\vvx(\tau)$. It is remarkable, that (\ref{eq:meinW}), in contrast to (\ref{eq:bdW}),
cannot be 
written as a function on Minkowski space because it depends on the four-velocity 
$\dot\vvx$. It can easily be seen, that (\ref{eq:bdW}) results from (\ref{eq:meinW})
by inserting 
$\dot\vvx(\tau)=(1,0,0,0)$, which, of course, is inconsistent because the detector is 
not at rest but moving along the trajectory $\vvx=\vvx(\tau)$. As a first test of our 
new correlation function, we insert all the possible stationary trajectories in 
Minkowski space, which are classified in \cite{let} into six groups. It turns out that 
in all six
cases the correlation function depends only on $\tau-\tau'$ which has to be expected
for a stationary movement. In contrast, inserting into (\ref{eq:bdW}) leads to a result 
which is invariant under time translation 
only in the cases of the detector at rest and in circular movement. We see this as 
further evidence that (\ref{eq:bdW}) is not suitable for the calculation of the 
detector response
and that (\ref{eq:meinW}) should be used instead. Furthermore, (\ref{eq:meinW}) 
is manifestly invariant under 
Lorentz transformation of the trajectory, which we consider as another advantage over
(\ref{eq:bdW}). For example, inserting $t(\tau)=\gamma\tau,\;\vx(\tau)=\gamma\vv\tau$ 
with an
arbitrary $\vv$ satisfying $\vv^2<1$ leads to the same correlation function 
(\ref{eq:ruhe}) as 
in the case of a detector at rest.

Concluding: Our proposal for calculating the transition rate at time $\tau$ for a 
detector moving on $\vvx(\tau)$ is
\begin{eqnarray}
\Fd=2\lime\int_0^{\infty}\d s\;\Re\;\bigl( e^{-i\omega s}\vev{\phi(\tau)
\phi(\tau-s)}\bigr)\quad,
\end{eqnarray}
where
\begin{eqnarray}
\cav\phi(\tau)\phi(\tau')\vac=\frac{1/4\pi^2}{\bigl( 
\vvx(\tau)-\vvx(\tau') -i\eps({\dot\vvx}(\tau)+
{\dot\vvx}(\tau'))\bigr)^2}\quad.
\end{eqnarray}
Thus, in order to calculate the time-dependent response of a detector in a manifestly 
causal way, i.e.\ by using only information on its movements in the {\em past}, we use 
a different integration formula and a different correlation function.

It is now easy to show that in the case of an uniformly accelerated trajectory
the correlation function is
\begin{eqnarray*}
\cav\phi(\tau)\phi(\tau')\vac=\frac{-1/16\pi^2}{\left(\sinh\left(\frac{\tau-\tau'}{2}
\right)-i\eps\cosh\left(\frac{\tau-\tau'}{2}\right)\right)^2}\quad,
\end{eqnarray*}
which clearly is invariant under time translations, as has to be expected from the
stationarity of the trajectory. We can thus write the detector response with an 
integration over the whole real axis, according to (\ref{eq:ttinv}):
\begin{eqnarray*}
\Fd=-\frac{1}{16\pi^2}\lime\int_{-\infty}^\infty\d s\;\frac{e^{-i\omega s}}
{\left(\sinh\left(\frac{s}{2}\right)-i\eps\cosh\left(\frac{s}{2}\right)\right)^2}
\end{eqnarray*}
and evaluate the integral with the help of contour-integration: We integrate along the 
real axis in positive direction and along the line $\Im s=2\pi$ in negative direction.
The only singularity within this contour lies at $s=2i\arctan\eps$.
A short calculation shows that because of the periodicity of $\cosh$ and $\sinh$, 
the value $I_\eps(\omega)$ of the integral is connected to the value of the residue by
\begin{eqnarray*}
2\pi i\Res\left(\frac{e^{-i\omega s}}
{\left(\sinh\left(\frac{s}{2}\right)-i\eps\cosh\left(\frac{s}{2}\right)\right)^2}
\;,\;s=2i\arctan\eps\right)=I_\eps(\omega)-e^{2\pi\omega}I_\eps(\omega)\quad.
\end{eqnarray*}
The value of the residue is found to be $-4i\omega e^{2\omega\arctan\eps}/1+\eps^2$
which leads to
\begin{eqnarray*}
I_\eps(\omega)=\frac{8\pi\omega}{1-e^{2\pi\omega}}\frac{e^{2\omega\arctan\eps}}
{1+\eps^2}
\end{eqnarray*}
and finally
\begin{eqnarray*}
\Fd=\frac{1}{2\pi}\frac{\omega}{e^{2\pi\omega}-1}\quad.
\end{eqnarray*}
This is of course the well-known thermal Unruh spectrum, which we derived here in a 
different way as usual, namely in a time-dependent, causal manner and by using a 
physically founded regularization of the correlation function. This approach may be 
considered more satisfying than the usual one.

\section{Application to a non-stationary example}

\begin{figure}[htbp]
\begin{center}
\psfrag{om}[][][0.7]{$\omega$}
\psfrag{Fd}[][][0.7]{$\Fd$}
\includegraphics[width=11cm]{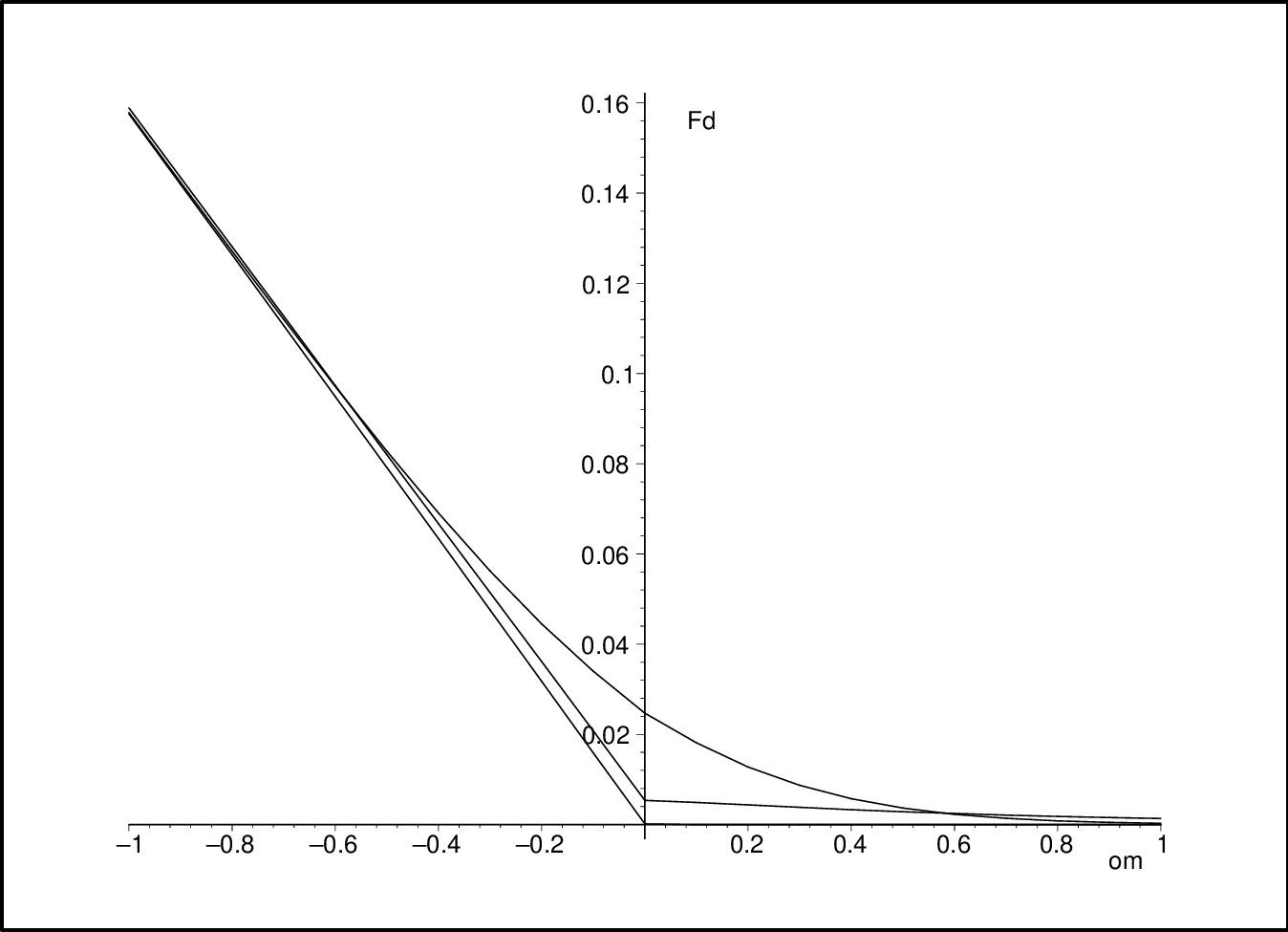}
\caption{Non-stationary Unruh-like spectrum}\label{huspe}
\end{center}
\end{figure}

Let us consider the non-stationary trajectory
\begin{eqnarray}
\label{eq:nonstat}
t(\tau)&=&\tau-\ln2+\sqrt{1+\frac{1}{4}e^{2\tau}}-
\ln\left(1+\sqrt{1+\frac{1}{4}e^{2\tau}}\right)\nonumber\\
x(\tau)&=&\frac{1}{2}e^\tau\quad,\quad y(\tau)\;=\;0\quad,\quad z(\tau)\;=\;0
\quad.
\end{eqnarray}
This movement interpolates smoothly between rest at $x=0$ and uniform acceleration along
$x^2-t^2=1$. Actually, it is easy to show that $t(\tau)\to\sinh\tau$ and 
$x(\tau)\to\cosh\tau$ if $\tau\to +\infty$. For $\tau\to -\infty$ we have instead
$t(\tau)\to \tau+1-2\ln2$ and $x(\tau)\to0$. The square of the four-acceleration
\begin{eqnarray*}
\vvb^2=-\ddot{t}^2+\ddot{x}^2=\frac{1}{1+4e^{-2\tau}}
\end{eqnarray*}
shows that the proper acceleration $\alpha=\sqrt{\vvb^2}$ is indeed smoothly increasing
from $0$ to $1$. The radiation spectrum is now time-dependent, of course. The result
of a numerical calculation for three times $\tau=-4,1,6$ is shown in figure \ref{huspe}.
At time $\tau=-4$ the movement has not really yet begun, so one gets the spectrum of the
detector at rest. At time $\tau=6$ the movement ran a fairly long time with 
approximately constant acceleration $\alpha=1$, so the thermal spectrum results, which
does not change any more at later times. It is remarkable, that at the intermediate time
$\tau=1$, i.e.\ $\alpha=0.65$, the transition-rate decays {\em slower} for high $\omega$
than in the case of the thermal spectrum at late times. This means, that the increasing
of acceleration takes place smoothly but not adiabatically: The radiation spectrum
is of a non-thermal nature, and not simply thermal with only a time-varying temperature.
This shows that not only acceleration contributes to the radiation effect,
but higher time derivatives as well.

\section{Conclusion}

Starting with the intention to derive an explicitly causal formulation for the 
transition rate of a non-inertial Unruh-DeWitt detector, we were led to the question
of which regularization of the scalar field's correlation function is appropriate to 
this problem. The usual prescription $t\to t-i\eps$ or the equivalent insertion of the 
convergence factor $e^{-\eps|\vk|}$ into the Fourier representation was shown to be 
inappropriate by a numerical calculation, which lead to a time-dependent unphysical
result in the case of uniformly accelerated motion. Following Takagi
we arrived at the correct regularization by considering the 
pointlike detector as the limit of a finite, rigid detector - where rigidity is
defined with respect to its own momentary rest system. The result is
\begin{eqnarray*}
\Fd=2\lime\int_0^{\infty}\d s\;\Re\;\bigl( e^{-i\omega s}\cav\phi(\tau)
\phi(\tau-s)\vac\bigr)\quad,\\
\cav\phi(\tau)\phi(\tau')\vac=\frac{1/4\pi^2}{\bigl( 
\vvx(\tau)-\vvx(\tau') -i\eps({\dot\vvx}(\tau)+
{\dot\vvx}(\tau'))\bigr)^2}\quad.
\end{eqnarray*}
We consider this as the physically well-founded form of the response function of an 
infinitesimal detector.

\bigskip

Finally, a short criticism of an often heared argument regarding the Unruh effect 
may be formulated. The fact that Unruh radiation is exactly {\em thermal} is 
astonishing and demands for some deeper explanation. A well-known argument is based on the
existence of an acceleration horizon which seperates the detector from some degrees
of freedom of the scalar field. Indeed, the eternal uniformly accelerating detector
or observer is causally separated from the ``left half'' of Minkowski space   by the 
horizon $t=x$.

Therefore, as the argument goes, the observer describes the global vacuum state 
$\vac$ of the 
scalar field as a density matrix which results from tracing out the degrees of freedom 
behind the horizon. This density matrix turns out to be thermal, with a temperature of 
$T=\alpha/2\pi$, cf.\ \cite{wald}. In this way the ocurrence of the thermal spectrum 
can be
understood as a consequence of the existence of an acceleration horizon and the 
associated ``hiding'' of degrees of freedom. 
This may look convincing, but is not without problem if considered from a causal 
viewpoint: On the one hand, the existence of an acceleration horizon can be stated
only if the {\em whole course} of the trajectory is known, in particular the future
behaviour up to $t=+\infty$. The existence of a thermal spectrum, on the other hand, 
is a fact which results only from the {\em past} movement of the detector. Then the
problem is posed, that an observer, who percives thermal radiation here and now,
{\em cannot} explain this by showing up the acceleration horizon, because this is not
an observable border, existing somewhere in spacetime, but rather an abstract object
which can be constructed only ``posthumously'' after the entire movement is known.
At any finite time, there exist for the accelerated observer no more and no less 
unobservable degrees of freedom as for the observer at rest.

The problem which - in my opinion - really calls for an explanation, is the fact,
that an observer moving along trajectory (\ref{eq:nonstat}), which smoothly goes over 
from rest to 
an arbitrarily long period of almost uniform acceleration, sees a radiation spectrum
which is thermal to any arbitrary precision - without making any assumptions on 
the future movement of the detector. Of course, the trajectory (\ref{eq:nonstat}) 
is defined for all
$t$, and with it comes the acceleration horizon $t=x$, but this does not enter the 
calculation. 
The spectrum measured at time $\tau$ would not change at all if the detector returns
to rest afterwards. Thus we have the following situation: A detector which is 
asymptotically at rest for $t\to\pm\infty$, is moving for an arbitrarily long 
(but finite) time with almost uniform acceleration and percives an (almost) thermal
radiation-spectrum. Because he returns from asymptotic rest to asymptotic rest, there
is {\em no} acceleration horizon in this case, and {\em no} hidden degree of freedom.
The question for a simple explanation of the (arbitrarily precise) thermality of the
observed radiation is but as important in this case as in the idealized case of
eternal hyperbolic movement.

\end{document}